\documentclass[useAMS,usenatbib,usegraphicx]{mn2e}

\DeclareMathAlphabet{\mathsc}{OT1}{cmr}{m}{sc}
\def\testbx{bx}%
\DeclareRobustCommand{\ion}[2]{%
\relax\ifmmode
\ifx\testbx\f@series
{\mathbf{#1\,\mathsc{#2}}}\else
{\mathrm{#1\,\mathsc{#2}}}\fi
\else\textup{#1\,{\mdseries\textsc{#2}}}%
\fi}

\title[Effect of the interactions and environment on nuclear activity]{Effect of
the interactions
and environment on nuclear activity}
\author[J. Sabater, P.~N. Best and M. Argudo-Fern\'andez]{J.
Sabater$^{1,2}$\thanks{E-mail:
jsm@roe.ac.uk (JS); pnb@roe.ac.uk (PNB); margudo@iaa.es (MAF)}, 
P.~N. Best$^{1}$\footnotemark[1]
and M. Argudo-Fern\'andez$^{2}$\footnotemark[1]\thanks{Full
table~\ref{table:env_par} is available
in electronic form at the CDS web site
http://cdsweb.u-strasbg.fr and in the online version of the journal.}\\
$^{1}$Institute for Astronomy (IfA), University of Edinburgh, Royal Observatory,
Blackford Hill, EH9 3HJ Edinburgh, U.K.\\
$^{2}$Instituto de Astrof\'{\i}sica de Andaluc\'{\i}a, CSIC,
              Apdo. 3004, 18080
              Granada, Spain}
\begin{document}

\date{Accepted XXXX Month XX. Received XXXX Month XX; in original form
XXXX Month XX}

\pagerange{\pageref{firstpage}--\pageref{lastpage}} \pubyear{2012}

\maketitle

\label{firstpage}

\begin{abstract}
We present a study of the prevalence of optical and radio nuclear activity with
respect to the environment and interactions in a sample of SDSS galaxies. The
aim is to determine the independent effects of distinct aspects of source
environment on the triggering of different types of nuclear activity. We defined
a local density parameter and a tidal forces estimator and used a cluster
richness estimator from the literature to trace different aspects of environment
and interaction. The possible correlations between the environmental parameters
were removed using a principal component analysis. By far the strongest trend
found for the active galactic nuclei (AGN) fractions, of all AGN types, is with
galaxy mass. We therefore applied a stratified statistical method that takes
into account the effect of possible confounding factors like the galaxy mass. We
found that (at fixed mass) the prevalence of optical AGN is a
factor 2--3 lower in the densest environments, but increases by a factor of 
$\sim 2$ in the presence of strong one-on-one interactions. These effects are
even more pronounced for star-forming nuclei. The importance of galaxy
interactions decreases from star-forming nuclei to Seyferts to LINERS to passive
galaxies, in accordance with previous suggestions of an evolutionary
time-sequence. The fraction of radio AGN increases very strongly (by nearly an
order of magnitude) towards denser environments, and is also enhanced by galaxy
interactions. Overall, the results agree with a scenario in which the mechanisms
of accretion into the black hole are determined by the presence and nature of a
supply of gas, which in turn is controlled by the local density of galaxies and
their interactions. A plentiful cold gas supply is required to trigger
star-formation, optical AGN and radiatively-efficient radio AGN. This is less
common in the cold-gas-poor environments of groups and clusters, but is enhanced
by one-on-one interactions which result in the flow of gas into nuclear regions;
these two factors compete against each other. In the denser environments where
cold gas is rare, cooling hot gas can supply the nucleus at a sufficient rate to
fuel low-luminosity radiatively-inefficient radio AGN. However, the increased
prevalence of these AGN in interacting galaxies suggests that this is not the
only mechanism by which radiatively-inefficient AGN can be triggered.
\end{abstract}


\section{Introduction}

Active galactic nuclei (AGN) are closely related to galaxy formation and
evolution. The black holes that power AGN are found in all massive galaxies and
their masses are tightly correlated with both the masses and the velocity
dispersions of the stellar bulges
\citep[e.g.][]{Marconi2003,Ferrarese2000,Gebhardt2000}. AGN may play an
important role in the feedback mechanisms that control the growth of massive
galaxies \citep[see review by][and references therein]{Cattaneo2009}.
Interaction between galaxies can drive gas into the central region of the galaxy
and trigger the AGN \citep{Shlosman1990,Barnes1991,Haan2009,Liu2011}. However, a
high fraction of AGN may by fueled by secular processes
\citep[e.g.][]{Silverman2011,Sabater2012}.

The effect of the large-scale environment and of galaxy interactions on the
triggering of an AGN has previously been studied by many authors, but
contradictory results have been found. Quasars are thought to be associated with
galaxy interactions \citep{Sanders1988,Urrutia2008,Letawe2010} or a higher
number of companions \citep{Hutchings1983,Bahcall1997,Serber2006}. Several
studies found a direct relation between interaction and the presence of an
optical AGN \citep[][]{Petrosian1982,Koulouridis2006,
Alonso2007,Rogers2009,Ellison2011,Liu2012,Hwang2012}. On the other hand, other
studies did not find this relation
\citep[][]{Bushouse1986,Schmitt2001a,Ellison2008,Li2008,
Darg2010,Slavcheva-Mihova2011}. In some cases, a lower prevalence of AGN is
found in environments with higher density of galaxies
\citep{Carter2001,Miller2003,Kauffmann2004}, or the AGN prevalence is found to
depend upon distance to the centre of the nearest group or cluster
\citep{vonderLinden2010}. In contrast, \citet{Alonso2012arxiv} found an increase
of the fraction of powerful AGN toward denser environments in a sample face-on
spiral galaxies. Some studies find signatures of mergers in active galaxies
\citep{Comerford2009,Smirnova2010, VillarMartin2011,VillarMartin2012}, but,
\citet{Grogin2005} found no connection between recent galaxy mergers and AGN.
Making use of radio observations, \citet{Kuo2008} and \citet{Tang2008} measured
the disruption of the atomic gas distribution and kinematics to trace
interactions and found a clear relation with the presence of an AGN.

With respect to X-ray selected AGN, a higher incidence of signs of disruption in
AGN \citep{Koss2010} and a higher fraction of AGN in pairs \citep{Silverman2011}
and in the centre of clusters \citep{Ruderman2005} have been found. However, a
relation between interaction and AGN was not found by \citet{Pierce2007},
\citet{Georgakakis2009} or \citet{Gabor2009}. \citet{Waskett2005} found that the
environments of AGN are indistinguishable from those of normal, inactive
galaxies. Finally, \citet{Silverman2009} found no difference between the
fraction of AGN in groups and in field galaxies. 

If radio selected AGN are considered, \citet{Best2005b} found a tendency for
radio-loud AGN to be located in richer environments. \citet{Reviglio2006}
confirmed this trend for radio AGN, which is in contrast to that widely found
for optical AGN. Brightest group and cluster galaxies are more likely to host a
radio-loud AGN than other galaxies with similar stellar masses
\citep[e.g.][]{Best2007}. \citet{Domingue2005} found an excess of radio AGN in
pairs of galaxies while no radio AGN were found in isolated galaxies
\citep{Sabater2008,Sabater2010p}. While \citet{Heckman1986} found many radio AGN
to be ``strongly peculiar'' in optical morphology, \citet{Tal2009} found no
correlation between radio AGN and tidal distortions. Nevertheless, there is
recent evidence of the triggering of powerful radio galaxies by interactions
\citep{RamosAlmeida2012,Bessiere2012arXiv}.

Part of the confusion from the results discussed above may involve the different
definitions of ``environment'' and ``interaction'' used in the previous studies.
This may be: 
a) a higher local density of neighbours,
b) the membership of a cluster or group versus field or
filaments,
c) relative location within a group or cluster, or,  
d) one-on-one interactions. 
These definitions can lead to contradictory results
since they trace different aspects of interaction and environment. It is known
that many interacting galaxy pairs can be found in low density regions (even
void; \citeauthor{Verley2007b} 2007; Argudo-Fern\'andez et al. in prep.),
consequently, the concept ``field'' is not equivalent to ``isolated''. We might
expect the amount and physical conditions of the gas in galaxies that is used to
fuel the AGN to be related to large-scale environment
\citep[e.g.][]{Sijacki2007}. On the other hand, interactions at the one-on-one
level are expected to set the gravitational disturbances that could trigger the
AGN. Therefore, it is fundamental to separate those different aspects when
studying their effect on nuclear activity.

A further complication is the type, and luminosity, of the AGN selected in
different studies. Although the emission of an AGN spans a wide range of the
electromagnetic spectrum \citep{Elvis1994}, the different methods and
wavelengths used to select AGN will also lead to the selection of different
kinds of AGN. All of the selection methods are affected by selection effects
\citep{Ho2008}, for example, X-rays have been used to detect AGN that did not
present optical emission lines \citep{Martini2002}. Selection based on optical
emission lines can discern between Seyfert and low-ionization nuclear
emission-line region \citep[LINER;][]{Heckman1980} types, however, the final
classification may depend on the exact processing method used and the
classification criteria applied \citep[e.g.][]{Kewley2006,Constantin2006}.
\citet{Best2005b} found that the probability of galaxy harbouring a radio AGN is
independent of its optical classification and conclude that low-luminosity radio
AGN and emission-line AGN are powered by different physical processes. Two main
feeding mechanisms can be found in AGN: a) ``standard'' high-excitation
radiatively-efficient mode, and, b) low-excitation radiatively-inefficient mode
\citep[see][and references therein]{Best2012}. Radiatively-efficient AGN include
luminous radio AGN (also called High Excitation Radio Galaxies or HERG) and the
optical or X-ray AGN while radiatively-inefficient ones are observed as
low-luminosity radio AGN \citep[Low Excitation Radio Galaxies or
LERG;][]{Hardcastle2007}. These types of feeding mechanisms can evolve in time
\citep{Hlavacek-Larrondo2012arxiv} and may be fundamental to explain the decline
of star formation in large elliptical galaxies \citep{Cattaneo2009}.

It is also important to consider that many properties of galaxies depend on the
environment, with clusters typically hosting more massive and more early-type
galaxies than the field \citep[e.g.][]{Blanton2009,Deng2009,Park2009,Deng2011}.
If samples are not well-matched (as has been the case for many previous studies)
then these differences can give rise to false dependencies. In particular it is
known that galaxy mass is one of the most (if not \textit{the} most) important
driving factors for the prevalence of: a) optical \citep[e.g.][]{Kauffmann2003},
b) radio \citep{Best2005b}, and c) X-ray \citep{Silverman2009,Tasse2011}
selected AGN. The prevalence of AGN also depends upon galaxy morphology
\citep{Moles1995,Schawinski2010,Sabater2012}. These properties are also closely
related to the environment via the density-morphology and density-luminosity
relations
\citep{Dressler1980,Kauffmann2004,Blanton2005,Deng2009,Park2009,Deng2011}.
Indeed, \citet{Park2009} suggest that galaxy properties are mainly driven by
mass and morphology, and environment plays only a secondary role. Hence, these
confounding factors should be carefully taken into account in order to obtain
unbiased results.

We aim to study independently the effects of the large-scale environment and the
smaller-scale interactions on the triggering of both radio and optically
selected AGN. In Sect.~\ref{sec:sample}, the sample and the data used are
presented. We define a set of parameters to quantify the environment and the
interactions that affect our galaxies, and these are described in
Sect.~\ref{sec:env_params}. The results of the statistical studies that estimate
the relation between the environment and interaction with the prevalence of
different types of active nuclei are presented in Sect.~\ref{sec:results}.
Finally, these results are discussed in Sect.~\ref{sec:dc} and the final
conclusions are presented Sect.~\ref{sec:conc}. \label{sec:cosmo} Throughout the
paper, the following cosmological parameters are assumed:
$\Omega_{\mathrm{m}}=0.3$, $\Omega_{\mathrm{\Lambda}}=0.7$ and $H_{\mathrm{0}}=
70\,\mathrm{km}\,\mathrm{s^{-1}}\,\mathrm{Mpc^{-1}}$.


\section{The sample and the data}
\label{sec:sample}

\subsection{The sample}

The sample was based on the seventh data release \citep[DR7][]{Abazajian2009} of
the Sloan Digital Sky Survey \citep[SDSS; ][]{York2000}. The base sample is
composed of galaxies from the main spectroscopic sample \citep[magnitudes
between $14.5 < r < 17.77$;][]{Strauss2002} with a redshift between 0.03 and
0.1. Galaxies with a redshift below 0.03 were discarded, a) to limit the search
for companions to a reasonable area of the sky around the target galaxies, b) to
lessen the possible effect of the incompleteness of galaxies with high
brightness and, c) to avoid the possible errors in the measurements of the
photometry due to the large size of the galaxies. Galaxies with a redshift above
0.1 were also discarded, a) to allow a good signal-to-noise for the emission
lines used for the optical activity classification and for the radio-continuum
emission used in the radio activity classification, and, b) to maximize the
number of companions with similar luminosities that are covered by the
spectroscopic survey. The evolutionary time between redshift 0.03 and 0.1 is
$\approx 9 \times 10^{8}$ years We take no account of any evolution of the
nuclear activity prevalence during this period, but this is supposed to be small
and in any case any evolutionary effects are minimized by the use of the
stratified statistical study. The sample is composed of galaxies from the table
\texttt{Galaxy} which have an assigned spectroscopic object \texttt{specObjID}
with a spectroscopic redshift (field \texttt{bsz}) within the limits.

A small percentage of luminous galaxies are not included in the main
spectroscopic survey at lower redshifts (those with $r < 14.5$ which at $z =
0.03$ corresponds to $L_{r} \sim 10^{10.2}\, \mathrm{L_{\odot}}$). At $z = 0.1$
the magnitude limit corresponds to a luminosity limit in \textit{r}-band of
$\sim 10^{9.4}\, \mathrm{L_{\odot}}$. For most of our analyses, stellar mass
(see next Section) limits of $10 \leq \log(M/\mathrm{M_{\odot}}) \leq 12$ were
applied to reduce the incompleteness. In any case, in our study we compute
relative fractions of galaxies instead of absolute numbers. These fractions
should not be biased by the small number of missing galaxies, especially since
we observe no significant trends of any of the AGN fractions with redshift
across our samples. The use of relative fractions, the cut in masses, and the
use of stratified statistical methods that take into account the mass of
galaxies should mitigate the possible effect of the incompleteness.

The environmental parameters that we aimed to obtain could be affected if the
target galaxy is close to the border of the SDSS footprint or near to a bright
star. The potential loss of a significant fraction of companion galaxies could
then cause an underestimation of the parameters. Hence, we discarded from our
study galaxies that could be affected by these problems. To do this we used the
New York University (NYU) Value-Added Galaxy Catalog \citep{Blanton2005b}, which
provides additional data to the SDSS sample. The parameter \texttt{NEXPECT}
found in the \texttt{simpleden} table gives an estimation of the number of
tracer galaxies that can be found within a given fixed radius of a target
galaxy. These tracer galaxies are homogeneously and randomly distributed but
only within the coverage region of the SDSS spectroscopic survey. Hence, if the
number is lower than a given value, it indicates that the region around the
target galaxy is not completely covered by the survey. We discarded galaxies
with $\mathtt{NEXPECT} < 0.45$. In some cases, there were no available data from
the NYU catalogue and we took the value of \texttt{NEXPECT} from the closest
galaxy with data within a maximum of 30 arcmin radius.  A visual inspection
\citep[using the software \textsc{TOPCAT};][]{TOPCAT} showed that this process
was reliable for the selection of galaxies in well-sampled SDSS areas. Four
galaxies were manually added and three regions close to the borders (126
galaxies in total) were manually discarded. Two additional regions containing
1594 galaxies close to the borders were flagged later after finding systematic
offsets for those galaxies in one of the environmental parameter that we
computed (the tidal estimator; Sect.~\ref{subsec:tidal}).

The final sample was composed of 267977 galaxies. Reliable estimations of 
the cluster richness using a friends of friends algorithm 
(as explained in Sect.~\ref{subsec:n}) could only be obtained for
galaxies brighter than $M_{r} - 5\log(\mathrm{h}) = -20$. These galaxies 
constitute the \textit{reduced sample} which was used when the 
cluster richness parameter was needed for the analysis. 
The total number of galaxies in each sample is shown 
in Table~\ref{table:numbers}.

\begin{table}
 \centering
 \begin{minipage}{80mm}
  \caption{Number of galaxies of each type.}
  \label{table:numbers}
  \begin{tabular}{lrrrr}
\hline
 Type\footnote{Meaning of the different types: Passive - galaxies with
$L_{[\ion{O}{iii}]} < 10^{6.5}$ L$_{\odot}$; Optical AGN - galaxies with
$L_{[\ion{O}{iii}]} \geq 10^{6.5}$ L$_{\odot}$ classified as Low Ionization
Nuclear Emission Regions (LINER), Seyferts or Transition Objects (TO); SFN -
Star Forming Nuclei galaxies with $L_{[\ion{O}{iii}]} \geq 10^{6.5}$
L$_{\odot}$; Radio AGN - galaxies with $L_{1.4\mathrm{GHz}} \geq 10^{23}$ W
m$^{-2}$ Hz$^{-1}$ classified as radio AGN including High Excitation Radio
Galaxies (HERG) and Low Excitation Radio Galaxies (LERG).}
 & Whole & Reduced & Whole & Reduced\\
  & sample & sample & sample & sample \\
  &   &  & $M$ lim.\footnote{Mass limited to
$10 \leq \log(M/\mathrm{M_{\odot}}) \leq 12$}
  & $M$ lim. \\
\hline
\hline
Total & 267977 & 107176 & 201425 & 106503 \\
\hline
~~Passive & 99180 & 47333 & 90174 & 47311 \\
~~Optical AGN & 57955 & 34266 & 55864 & 34255 \\
~~~~LINER & 16288 & 12315 & 16144 & 12313 \\
~~~~Seyfert & 7170 & 3691 & 6763 & 3689 \\
~~~~TO & 34497 & 18260 & 32957 & 18253 \\
~~SFN & 110842 & 25577 & 55387 & 24937 \\
\hline
~~Radio AGN & 1137 & 1015 & 1129 & 1015 \\
~~~~LERG & 1087 & 979 & 1082 & 979 \\
~~~~HERG & 47 & 34 & 44 & 34 \\
\hline
\end{tabular}
\end{minipage}
\end{table}

\subsection{AGN classification}

Our aim was to check the dependence of the different nuclear activity types on
the environment and interaction. For this study, we considered
optically and radio selected nuclear activity types.

Data based on optical spectra were drawn from the Max Plank Institute for
Astrophysics and Johns Hopkins University (MPA-JHU) added value spectroscopic
catalogue \citep[cf.][]{Brinchmann2004arxiv}. Information about total stellar
mass \citep{Kauffmann2003a} and the classification of the optical nuclear
activity based on the accurate measurement of emission-line fluxes can be found
in this catalogue. The classification of optical activity is based on the
typical diagnostic diagrams \citep[BPT diagrams;][]{Baldwin1981,Kauffmann2003}.
The following optically-selected nuclear activity types are distinguished: 
a) star forming nuclei (SFN), 
b) transition objects (TO; which may contain a mix of star-formation and an
AGN) 
c) Seyfert, 
d) LINER, and 
e) passive galaxies.
The types LINER, Seyfert and TO are aggregated and considered as the optical
AGN. It is important to notice that the catalogue does not include galaxies that
present broad emission lines as Seyfert 1. In practically all of the galaxies
(99.9 per cent) an [\ion{O}{iii}] emission line (at 500.7-nm) with a luminosity
of $10^{6.5}$ L$_{\odot}$ would be detectable above the noise in our sample. To
obtain a complete unbiased sample of active galaxies traced by optical emission
lines, we used this cutoff for galaxies classified as optical AGN or as star
forming (SF). The galaxies with [\ion{O}{iii}] luminosities below this threshold
were included in the passive galaxies group.

\citet{Best2012} obtained the radio nuclear activity classification for galaxies
in the MPA-JHU catalogue. The method followed the techniques presented in
\citet{Best2005a} and used radio-continuum data from the National Radio
Astronomy Observatory (NRAO) Very Large Array (VLA) Sky Survey
\citep[NVSS;][]{Condon1998} and the Faint Images of the Radio Sky at Twenty
centimetres \citep[FIRST;][]{Becker1995} databases. The separation between
``low-excitation'' and ``high-excitation'' radio galaxies (LERG and HERG
respectively) is also given. The radio-continuum luminosity limit at $z = 0.1$,
corresponding to the flux density level (5 mJy) achieved in the cross-matching
process, is $L_{1.4\mathrm{GHz}} \approx 10^{23}$ W m$^{-2}$ Hz$^{-1}$. We
consider as radio AGN galaxies those that are classified as a radio AGN by
\citet{Best2012} and have a radio luminosity above this threshold.

Table~\ref{table:numbers} shows a breakdown of the number of galaxies in each
sample depending on their classification.

\subsection{Potential companions}
\label{sec:companions}

For our tidal forces estimator (see Sect.~\ref{subsec:tidal}), when
investigating potential companions for our targets we need to consider not only
the SDSS spectroscopic galaxies but also galaxies with similar luminosities that
were not covered by the spectroscopic survey (e.g. due to fibre allocation
restrictions) and galaxies with lower luminosities (below the spectroscopic
magnitude limit) that could be companions of our targets. We considered the
objects in the SDSS Galaxy catalogue with: a) r magnitude below 22.0, and b)
spectroscopic redshift below 0.11, or photometric redshift minus three times its
error below 0.10. With this magnitude limit we include potential companions at
least $\approx 50$ less luminous than the least luminous target galaxy at a
given redshift. The spectroscopic redshift was used when available and the
photometric redshift was considered in the other cases (whenever it was used our
methods took into account its larger relative error). Various photometric
redshift catalogues have been constructed from SDSS data, and we compared these
against spectroscopic redshift catalogues in different magnitude ranges to
choose which to use. We selected the photometric redshift provided in the column
\texttt{photozcc2} of the schema \texttt{Photoz2} \citep{Oyaizu2008,Cunha2009}
for galaxies with a magnitude fainter than 17.77 because its distribution with
respect to the redshift was clearly the most accurate in the magnitude range
$17.77 < r < 22.0$. For galaxies without spectra and $r < 17.77$, we used the
photometric redshift provided in the column \texttt{z} of the schema
\texttt{Photoz}, which seemed to agree better with spectroscopic redshifts for
galaxies at these magnitudes.


\section{Environmental parameters}
\label{sec:env_params}

We chose three different parameters to trace different aspects of the
environment and the interaction level. We defined a local density and a
tidal forces estimator and obtained an estimation of group richness from the
literature.

\subsection{Density estimator}

The density estimator (also called local density estimator) traces the density
of companions around the target galaxy which is related to the dark matter halo
density and is usually defined by the distance to the 5th or 10th nearest
neighbour \citep[e.g. ][]{Dressler1980,Miller2003,Best2004,Verley2007b}. It can
be defined as $$\eta_{t} \equiv \log \left(\frac{k}{\mathrm{Vol}(r_{k})} \right)
= \log \left(\frac{3 k}{4 \pi r_{k}^{3}} \right),$$ where $r_{k}$ is the
projected-distance (in Mpc) to the $k^{\mathrm{th}}$ nearest neighbour. It is
possible to find different definitions of this parameter in the literature. In
some cases, a cylindrical volume or a surface density \citep[e.g.][]{Best2004}
is used. Note that the main factor that drives the parameter is the ratio
$k/r_{k}$ and these different definitions produce only a scaled or shifted
version of $\eta_{t}$.

We dynamically obtained a list of potential companions around each of our
galaxies within a given radius. We considered as companions galaxies with a
difference in spectroscopic redshift lower than 0.01 ($| z_{c} - z_{t} | \leq
0.01$) corresponding to $\Delta v < 3000$ km s$^{-1}$ and with luminosities in
\textit{r}-band between $10^{9.86}$ and $10^{12.0}$ L$_{\odot}$. The lower limit
of $10^{9.86}$ L$_{\odot}$ corresponds to the luminosity limit of the main
spectroscopic sample at redshift 0.1. Once selected, all the potential
companions were supposed to be at the same distance as the target galaxy and the
k-corrections were computed using the analytical method of
\citet{Chilingarian2010}. We used the distance to the 10th nearest neighbour.

\begin{figure}
\centering
 \includegraphics[width=7.5cm]{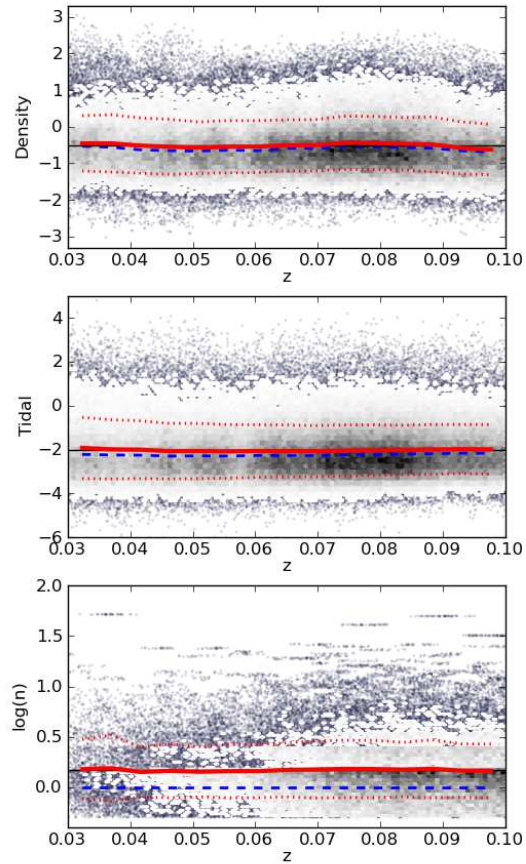}
  \caption{Dependence of the environmental parameters on the redshift. The
relations of the local density (upper panel), tidal forces (middle panel) and
cluster richness (lower panel) estimators with respect to the redshift are
shown. The values for the target galaxies are shown as a grey-scale density
cloud and dots for the outliers. Fifteen redshift bins where used. The mean (red
solid line), its error (red dotted line), and the median (blue dashed line) for
each redshift bin are plotted. The underlying black solid line marks the mean
value for the whole sample. A random offset (ranging from $-0.5$ to $0.5$) was
introduced in the cluster richness parameter ($n$) to allow a proper
visualization of the plot.}
  \label{fig:z_params}
\end{figure}

The main spectroscopic sample is $\sim$94 per cent complete \citep{Strauss2002}.
The spatial distribution of the few galaxies with similar magnitudes that are
not covered by the spectroscopic survey could affect the accuracy of the
environmental estimation for some galaxies. However, given the size of the
sample under study this is unlikely to be important unless it introduces
systematic biases. One indicator of a systematic bias in the local density is
the presence of a dependence with redshift. If the number of companions is
under- or over-estimated by our method for a given redshift range, a monotonic
trend in the mean values of the estimator should be expected. In
Fig.~\ref{fig:z_params} the dependence of the local density (upper panel) with
redshift is shown. There is no strong trend with redshift for the parameter. The
variation of the mean with redshift is less than 0.84 per cent with respect to
the variation range of the local density.

\subsection{Tidal forces estimator}
\label{subsec:tidal}

The tidal estimator traces the relative tidal forces exerted by companions
with respect to the internal binding forces of the target galaxy
\citep{Verley2007b}. The
definition is
$$Q_{t} \equiv 
\log \left( \sum_{i} \frac{M_{t}}{M_{i}} 
     \left( \frac{D_{t}}{d_{i,t}} \right)^{3} \right) = 
\log \left( \sum_{i} \frac{Lr_{t}}{Lr_{i}} 
     \left( \frac{2 R_{t}}{d_{i,t}} \right)^{3} \right),$$ 
where $D$ is the estimated diameter of the target galaxy
and $R$ is its radius, 
$d$ is the distance between the target and the companion,
$Lr$ is the corrected luminosity in \textit{r}-band and
$M$ is the mass of the galaxy. The luminosity in \textit{r}-band 
is used, in this case, as a proxy for the mass of the galaxy. To broadly match
the definition of diameter used in the literature \citep[projected major axis of
a galaxy at the 25 mag/arcsec$^2$ isophotal level or $D_{25}$;][]{Verley2007b},
we estimated the radius by using the Petrosian radius containing 90 per cent of
the total flux ($r_{90}$; provided in the SDSS catalogue) scaled by a factor
1.46 (see Argudo-Fern\'andez et al. in prep.)

For the tidal forces estimator, the inclusion of the photometric galaxies (see
Sect.~\ref{sec:companions}) is critical, both because the SDSS fibre allocation
process may not allow close companions to be targeted, and because the tidal
forces are often dominated by nearby galaxies of lower luminosity. The first
step before computing the tidal parameter was therefore to select which
potential companions were actually associated to our target galaxy rather than
being foreground/background galaxies. For each potential companion galaxy
without spectroscopy, we estimated the probability for it to be associated to
the target galaxy by chance, and we used this probability to discard spurious
companions. Specifically, for a hypothetical target at a redshift $z$, we
computed the sky density of galaxies within the SDSS survey with an
\textit{r}-band magnitude brighter than a limit ($r_{lim}$) and with the
difference between their photometric redshift and the redshift of the target
being less than $N$ times the error of the photometric redshift. Hence, we
obtained the surface density of galaxies as a function of $z$, $r_{lim}$ and
$N$. For each potential companion, the expected surface density of companions
with its properties ($r < r_{c}$, $z = z_{t}$ and $N < \frac{z_{p,c} -
z_{t}}{\Delta z_{p,c}}$) was computed (the sub-indexes $t$ and $c$ denote target
and companion respectively, $z_{p}$ is the photometric redshift and $\Delta
z_{p}$ is its error). If the expected number of companions within the area
defined by the distance between the companion and the target galaxy is less than
0.05, we considered the candidate galaxy as a genuine companion. Potential
companions with spectroscopy were selected if $| z_{c} - z_{t} | \leq 0.01$. The
selected companions (photometric and spectroscopic) were presumed to be at the
same distance as the target galaxy and k-corrections were applied.

The search radius for tidal companions was limited to 3 h$^{-1}$\,Mpc around the
target ($\sim 4.29$ Mpc using the cosmology especified in
Sect.~\ref{sec:cosmo}). We empirically checked that increasing the search radius
above this limit produced no appreciable change in the tidal parameter. If the
relative force (tidal force over internal binding force) exerted by an
individual companion is defined as $q_{t,i} =
({Lr_{t}}/{Lr_{i}})({2R_{t}}/{d_{i,t}})^{3}$ such that $Q_{t} = \log(\sum_i
q_{t,i})$,   the contribution of companions above the 3 h$^{-1}$\,Mpc limit was
always $q_{t,i} \leq 10^{-6}$, while the mean for closer companions is $q_{t,i}
\sim 10^{-2}$. After the rejection of spurious candidates, the tidal parameter
was computed. In the rare case that no companions were selected ($N = 684$), an
upper limit of $Q = -6$ was assigned to the galaxy. A galaxy $\approx 100$ times
more massive than the target would be required to produce this effect at a
distance of $\approx 3$ h$^{-1}$\,Mpc, our search radius limit.

As a test for systematic biases, the middle panel of Fig.~\ref{fig:z_params}
shows the dependence of the tidal forces estimator with the redshift. In this
case, the variation of the mean is less than 0.37 per cent of the variation
range of the parameter.

It is important to note that the tidal estimator uses the projected
two-dimensional distance as the distance between the target and the companion.
This will lead to a general overestimation of the tidal parameter with respect
to its actual value. The overestimation can be more severe in regions with a
high density of galaxies. However, as the projected distance depends only on the
(random) line of sight with respect to the observer, we do not expect a
systematic bias to be introduced by this effect. The expected effect is a
possible decrease in the strength of the correlations between the tidal
parameter and other properties, due to a degree of random scattering in the
tidal paremeter measurement.

\begin{figure*}
\centering
 \includegraphics[width=15cm]{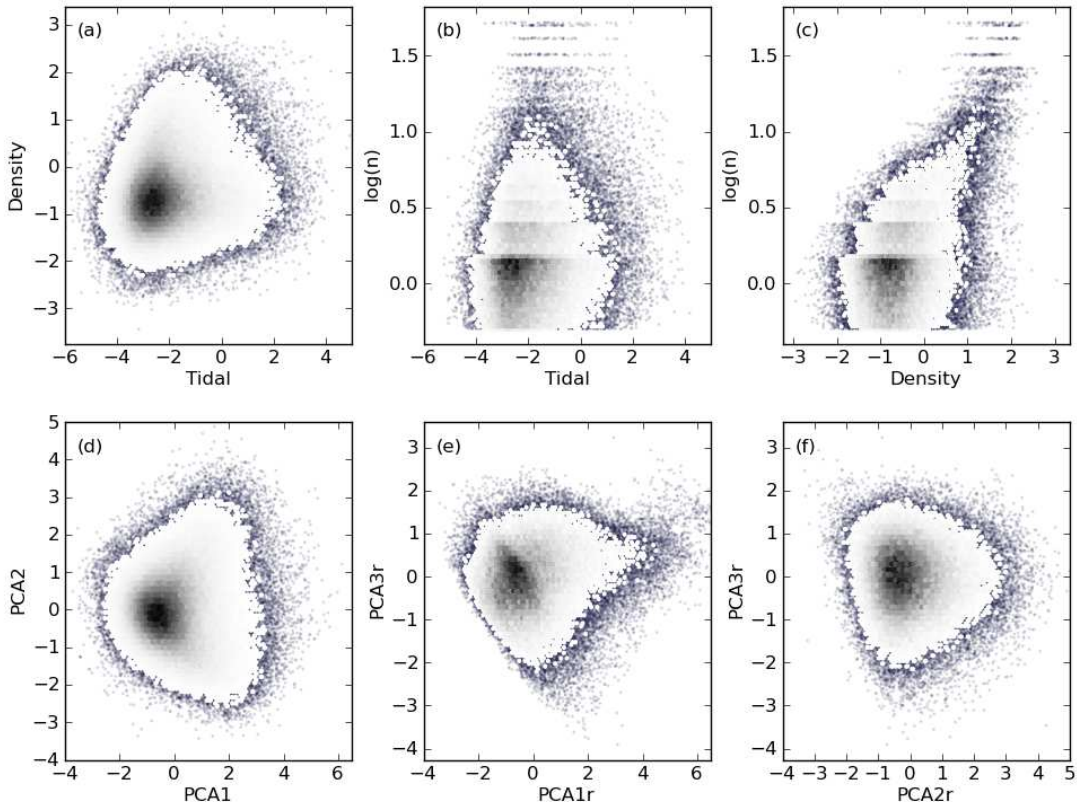}
  \caption{Environmental parameters. Panels \textbf{a} and \textbf{d} show the
distribution for the whole sample and the rest of the panels for the reduced
sample. The values for the target galaxies are shown as a grey-scale density
cloud and dots for the outliers. A random offset (ranging from $-0.5$ to $0.5$)
was introduced in the cluster richness parameter ($n$) to allow a proper
visualization of the plots.}
  \label{fig:params}
\end{figure*}

\subsection{Cluster richness estimator}
\label{subsec:n}

The last parameter is an estimation of the number of galaxies in the cluster or
group to which our target galaxy belongs. \citet{Tago2010} extracted groups of
galaxies from the SDSS DR7 using a modified friends-of-friends (FoF) algorithm.
A flux-limited and three volume-limited samples were provided. The groups of
galaxies were detected and a value ($n$) was assigned to each galaxy
corresponding to the number of galaxies in its own group, the cluster richness.
The number density of groups detected by FoF techniques depends strongly on
redshift for a flux-limited sample, hence, a volume-limited sample is required
to avoid any biases. The volume-limited sample which covers all our redshift
range is the DR7-20. The cutoff is $M_{r} - 5\log(\mathrm{h}) = -20$ ($M_{r}
\approx -19.2$) which covers from $z = 0.02595$ to $z = 0.11012$ given the
magnitude limits of the SDSS. Therefore, in our analyses, the \textit{reduced
sample} is composed of the galaxies that are bright enough in \textit{r}-band
and have an assigned value for the cluster richness. The value of the cluster
richness ranges from 1 to 52.

The dependence of cluster richness with redshift is shown in the lower panel of
Fig.~\ref{fig:z_params}. The variation of the mean is less than 3.16 per cent of
the variation range of the parameter. The presence of a few clusters with a high
number of galaxies cause this slightly higher value with respect to the other
parameters. The low values found for the variance of the mean of the three
environmental parameters with respect to the redshift make us consider that they
are not biased.

\subsection{Principal component analysis}

The three presented parameters provide three complementary measures of the
environment. The distribution of the local density and the tidal forces
estimator is shown in the panel \textbf{a} of Fig.~\ref{fig:params}. Galaxies
located at higher densities present medium to high tidal values but the highest
tidal values are found for galaxies located in regions with medium to low
densities. These galaxies are sometimes labeled as ``field'' galaxies according
to their local density but they are clearly interacting. Hence, it is important
to notice that the concept ``field'' and ``isolated'' are not equivalent. The
panels \textbf{b} and \textbf{c} of Fig.~\ref{fig:params} show the relation of
the cluster richness estimator with the local density and the tidal parameter
respectively. Each value of the cluster richness spans over a wide range of
tidal and density values. In the case of the local density, it is clear that
richer clusters present higher values of the local density.

We performed a Principal Component Analysis (PCA) to break the parameters down
into the independent physical processes by removing the possible correlations
between them. For example, part of the relation between the density and the
tidal parameter was expected to be produced by projection effects, with
generally higher tidal values for galaxies that are located in denser
environments. Some degree of relation between the group richness and the density
was also expected, as found. 

We obtained two PCA components by combining the local density and the tidal
parameters for the whole sample and three components for the reduced sample
where we added the group richness (the suffix ``r'' was added to distinguish
these principal components; we entered $\log(n)$ instead of $n$ as the input of
the PCA). The PCA components are shown in Table~\ref{table:PCA} and the
normalization values applied to the environmental parameters are shown in
Table~\ref{table:PCA_norm}. The effect of the two-dimensional projection on the
tidal parameter was at least partially mitigated after considering the
correction introduced by PCA2 with respect to the density (and cluster
richness).

\begin{table}
 \centering
 \begin{minipage}{80mm}
  \caption{Principal Component Analysis results.}
  \label{table:PCA}
  \begin{tabular}{cccccc}
\hline
 & & Density & Tidal & log($n$) & \textit{Var.}\footnote{Variance carried by
the PCA component.} \\
\hline
\hline
\textit{whole} & PCA1 & 0.707 & 0.707 & -- & 57\% \\
\textit{sample} & PCA2 & $-$0.707 & 0.707 & -- & 43\% \\
\hline
\textit{reduced} & PCA1r & 0.661 & 0.351 & 0.663 & 53\% \\
\textit{sample}  & PCA2r & $-$0.254 & 0.937 & $-$0.242 & 31\% \\
                 & PCA3r & $-$0.706 & $-$0.008 & 0.708 & 16\% \\
\hline
\end{tabular}
\end{minipage}
\end{table} 

As it is shown in Table~\ref{table:PCA}, PCA1 and PCA1r are composed of a
positive contribution of all the environmental parameters. PCA2 results from a
positive contribution of the tidal parameter and a negative contribution of the
density estimator, with PCA2r also having a negative cluster richness
contribution. This similarity leads us to consider that PCA1 and PCA1r trace
practically the same fundamental property, as do PCA2 and PCA2r. We later
checked that the results obtained using either PCA1 and PCA2 or PCA1r and PCA2r
were similar, hence, we will show the numbers obtained using the tidal, density,
PCA1 and PCA2 for the whole sample and the cluster richness and PCA3r for the
reduced sample. The relations between the PCA components are shown in
Fig.~\ref{fig:params}.

\begin{figure}
\centering
 \includegraphics[width=8cm]{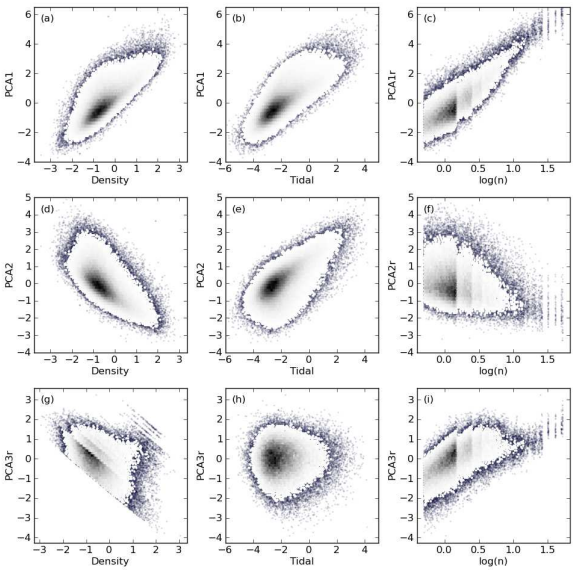}
  \caption{Relation between the environmental parameters and the PCA. Panels
\textbf{a}, \textbf{b}, \textbf{d} and \textbf{e} show the distribution for the
whole sample and the rest of the panels for the reduced sample. The values for
the target galaxies are shown as a grey-scale density cloud and dots for the
outliers. A random offset (ranging from $-0.5$ to $0.5$) was introduced in the
cluster richness parameter ($n$) to allow a proper visualization of the plots.
This random offset was also propagated to the PCA components that involve the
reduced sample for the same reason.}
  \label{fig:params_crossed}
\end{figure}

We interpret the PCA components as follows:
\begin{enumerate}
\item PCA1 and PCA1r follow the direction of the increase of the density (and
the cluster richness) and also the tidal forces. We consider
it to trace the overall interaction level and environmental nature of a
galaxy.
\item PCA2 and PCA2r are mainly driven by the difference between the density
(and cluster richness in the case of PCA2r) with respect to the tidal forces. A
higher value traces
higher one-on-one interactions while a lower value traces galaxies that are
relatively isolated for its overall environment.
\item PCA3r is practically not affected by the tidal force and is driven 
by the difference between the density and the cluster richness. In this 
case, a galaxy within a rich group but with low local density values has 
high PCA3r values and vice-versa. Hence, it would probably 
trace 
scatter in density versus cluster richness relation, being high where
densities are low for that particular $n$ (e.g. cluster outskirts) and low in
particularly locally-compact structure (e.g. compact groups).
\end{enumerate}
In Fig.~\ref{fig:params_crossed}, the relations between the PCA components and
the environmental parameters are shown. PCA parameters were computed applying
the values presented in Table~\ref{table:PCA} to the normalised version of the
environmental estimators obtained using the parameters presented in
Table~\ref{table:PCA_norm}. The final values of the environmental estimators for
all galaxies in the sample are presented in Table~\ref{table:env_par}. 

\begin{table}
 \centering
 \begin{minipage}{80mm}
  \caption{Principal Component Analysis normalization values.}
  \label{table:PCA_norm}
  \begin{tabular}{cccc}
\hline
 & Parameter & Mean & Standard deviation \\
\hline
\hline
\textit{whole} & Tidal & $-$2.0275 & 1.1952 \\
\textit{sample} & Density & $-$0.5082 & 0.7254 \\
\hline
\textit{reduced} & Tidal & $-$1.8951 & 1.1788 \\
\textit{sample}  & Density & $-$0.4865 & 0.7100  \\
                 & log($n$) & 0.1708 & 0.2748  \\
\hline
\end{tabular}
\end{minipage}
\end{table}

\begin{table*}
 \centering
 \begin{minipage}{160mm}
  \caption{Environmental parameters. The full version of the table can be found
in the online version of the journal.}
  \label{table:env_par}
  \begin{tabular}{rrrrrrrrrrrrc}
\hline
 plate & mjd & fiberID & Density & Tidal & log(n) & PCA1 & PCA2 & PCA1r 
 & PCA2r & PCA3r & Mass & Class.\footnote{Classification code. The first pair
of characters refer to radio activity and the second pair to optical activity.
First character, radio AGN classification: \texttt{R} Radio AGN; \texttt{-} Not
a radio AGN. Second character, type of radio AGN: \texttt{H} HERG; \texttt{L}
LERG; \texttt{-} Non classified. Third character, optical AGN classification:
\texttt{A} optical AGN; \texttt{S} star forming nuclei; \texttt{-} Passive.
Fourth character, type of optical AGN activity: \texttt{S} Seyfert; \texttt{L}
LINER; \texttt{T} Transition object; \texttt{-}
Not an optical AGN.}\\
\hline
\hline
 267 & 51608 &    1 & $-1.318$ & $-2.881$ & - & $-1.294$& $ 0.284$ & - & - & - &
 9.62 & \texttt{--S-} \\
 286 & 51999 &    1 & $-1.159$ & $-1.904$ & $ 0.000$ & $-0.562$& $ 0.708$ &
$-1.041$ & $ 0.384$ & $ 0.229$ & 10.63 & \texttt{--AT} \\
 291 & 51928 &    1 & $-0.497$ & $-2.878$ & - & $-0.492$& $-0.514$ & - & - & - &
10.25 & \texttt{----} \\
 297 & 51959 &    1 & $-0.943$ & $-2.546$ & - & $-0.730$& $ 0.116$ & - & - & - &
10.32 & \texttt{----} \\
 299 & 51671 &    1 & $-0.189$ & $-2.570$ & $ 0.000$ & $-0.010$& $-0.632$ &
$-0.336$ & $-0.491$ & $-0.731$ & 10.77 & \texttt{--AS} \\
 $\cdots$ & $\cdots$ & $\cdots$ & $\cdots$ & $\cdots$ & $\cdots$ 
 & $\cdots$ & $\cdots$ & $\cdots$ & $\cdots$ & $\cdots$ & $\cdots$ & $\cdots$\\
\hline
\end{tabular}
\end{minipage}
\end{table*} 


\section{Results}
\label{sec:results}

Our aim was to compute the relative fraction of each type of AGN with respect to
the environmental and PCA parameters. The strong dependence of the prevalence of
AGN with the mass of the host galaxy in the case of optical
\citep{Kauffmann2003} and radio \citep{Best2005b} AGN has to be considered.
Hence, to begin with, we divided the sample in four bins with masses between
$\log(M) = 10\,\mathrm{[M_{\odot}]}$ and $\log(M) = 12\,\mathrm{[M_{\odot}]}$
for radio AGN and six bins with masses between $\log(M) =
10\,\mathrm{[M_{\odot}]}$ and $\log(M) = 11.8\,\mathrm{[M_{\odot}]}$ for optical
AGN; the AGN fractions as a function of environmental parameters are shown for
each of these mass bins in Fig.~\ref{fig:mass_slices}. The monotonic increase of
the fraction of active galaxies with respect to the mass of the host is clearly
visible. This increase is steeper for radio AGN than for optical AGN. For radio
AGN the trends of AGN fraction with environmental parameters are very similar
for the four mass slices. In the case of optical AGN the trends are consistent
for the density, PCA1 and PCA2 but for the tidal estimator, the cluster richness
and PCA3r the results are not so clear. The bin of masses between $\log(M) =
11.5\,\mathrm{[M_{\odot}]}$ and $\log(M) = 11.8\,\mathrm{[M_{\odot}]}$ tends to
not to agree with the others, probably because of the low number of galaxies
within this mass range (389 galaxies in the reduced sample or 432 galaxies in
the whole sample which represent 0.4 or 0.2 per cent of the total number of
galaxies considered respectively).

\begin{figure*}
\vbox to220mm{\hfil 
\includegraphics[]{images/figure4s.epsi}
\hfil}
\label{fig:mass_slices}
\end{figure*}
\setcounter{figure}{4}

Although the overall trends could be inferred from Fig.~\ref{fig:mass_slices},
methods that allow an aggregation of the information of the different mass
slices and hence a quantitative measurement of the effect of the environmental
or PCA parameter are preferable. With the aggregation, a better signal to noise
ratio can be obtained and the quantification allows to draw more robust
conclusions. Therefore, we first defined an aggregated ratio that joins the
trends found in the different slices, and then, a statistical test that takes
into account the effect of the mass (as a possible confounding factor) was
applied.

\subsection{Stratified study}

To discard the effect of the mass on the fraction of AGN, we performed a
stratified study of the ratio of AGN with respect to each environmental
parameter. The sample was sliced in several mass strata (defined by the subindex
$i$) and also in several bins of the environmental parameter considered (defined
by the subindex $j$). This two-dimensional grid $(i,j)$ was fixed, and the
following values were counted for each bin: a) $n_{i,j}$; the total number of
galaxies in a bin and, b) $n_{a,i,j}$; the total number of galaxies with a given
type of nuclear classification in a bin (we will use the term AGN from now on,
although it can refer to any type of nuclear activity including passive
galaxies). The probability of a galaxy to harbour an AGN in each mass stratum is
$p_{a,i} = {\sum_{j} n_{a,i,j}}/{\sum_{j} n_{i,j}}$. If there is no
environmental dependence of the AGN fraction, then an estimated value of the
number of AGN in each bin of the grid can be computed as $n_{a\_est,i,j}=p_{a,i}
n_{i,j}$. This estimate accounts for the dependence on mass, so any remaining
effect is due to environment.\footnote{Notice that we remove largely the mass
effects but small residual trends can remain if e.g. AGN and all galaxies are
differently distributed within the defined mass bins.} Hence, a trend of the
prevalence of AGN with respect to the environmental parameter, that takes into
account the effect of the mass, can be extracted by comparing $n_{a\_est,i,j}$
and $n_{a,i,j}$. The general trend with respect to the environmental parameter
can be found by folding into the mass axis: $n_{a\_est,j} = \sum_{i}
n_{a\_est,i,j}$, $n_{a,j} = \sum_{i} n_{a,i,j}$ and $n_{j} = \sum_{i} n_{i,j}$.
Then, the probabilities of harbouring an AGN with respect to the environmental
parameter are $p_{a,j} = {n_{a,j}}/{n_{j}}$ while the estimated value (taking
account of the mass distribution) is $p_{a\_est,j} = {n_{a\_est,j}}/{n_{j}}$.
The errors on these values can be calculated following a binomial distribution.
The final trend, defined by the relative fraction between the observed and the
estimated probabilities, is $f_{i} = {p_{a,i}}/{p_{a\_est,i}}$. A value of
$f_{i} - \Delta f_{i} > 1$ or $f_{i} + \Delta f_{i} < 1$ means that, for this
range of the environmental parameter, there is a significant increase/decrease
of the observed fraction of AGN with respect to the expected one, after
correcting for the effect of the mass.

\begin{figure*}
\centering
 \includegraphics[width=15cm]{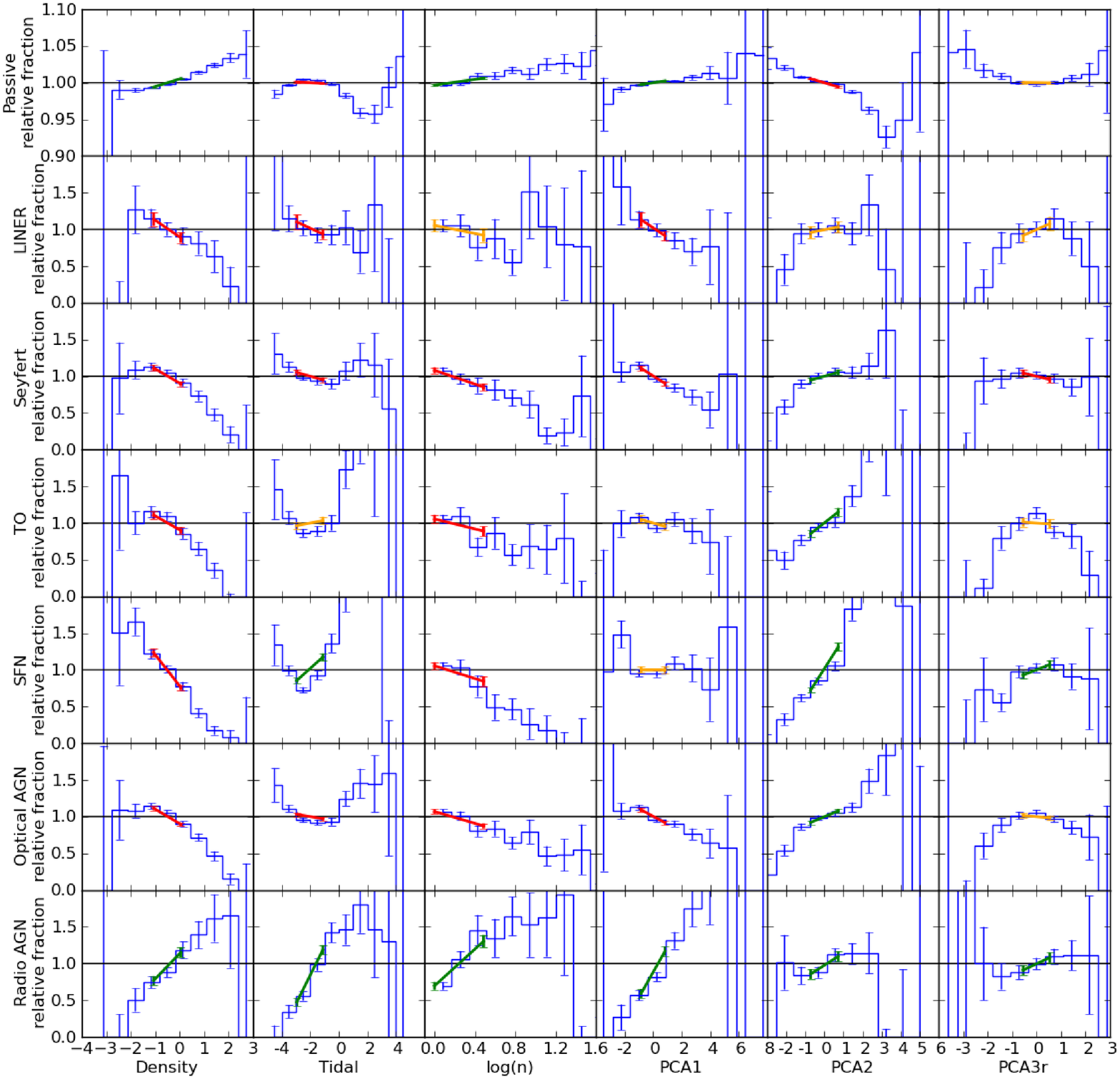}
  \caption{Relative fraction of galaxies of one type, corrected for the effect
of mass, with respect to the different environmental parameters. The blue thin
solid lines mark the detailed trend of $f$ with respect to each environmental
parameter. The thick solid lines defined by two points show the coarse general
trend of $f$. The colour of the thick colour line indicates whether a flat trend
is compatible within the error (orange) or not (increase - green; decrease -
red).}
  \label{fig:prev}
\end{figure*}

In Fig.~\ref{fig:prev}, $f_{i}$ is shown for the different types of nuclear
activity and environmental parameters. Seven bins were used in the mass axis and
10 bins in the environmental parameter axis defining a detailed trend. A coarse
general trend using two bins is also shown. These two bins were separated at the
median of each environmental or PCA parameter except in the case of the cluster
richness for which galaxies with $n=1$ were grouped in the first bin and the
rest in the second one. We also divided the cluster richness in two bins using
the median after discarding galaxies with $n=1$ and found that the trends were
the same.

The trends with respect to the density are clear and monotonic. There is an
increase with density in the prevalence of passive galaxies and radio AGN and a
decrease for all the optical AGN types and SFN. The trends are similar for the
cluster richness but slightly less acute. The tidal parameter shows a non-linear
trend for optically selected types with the lowest AGN fractions at intermediate
values. This may be related to correlations with other parameters. For example,
the galaxies which are located in high density regions tend to have medium tidal
values (see panel \textbf{a} in Fig.~\ref{fig:params}), and the strong trends
found with respect to the density might therefore affect the relative ratio for
galaxies with mid tidal values. This interpretation is supported by the fact
that there is a direct clear relation between PCA2 and the presence of all types
of activity; this also demonstrates the power of the PCA analysis in separating
the different physical processes. The trends with respect to PCA1 are similar to
the density ones but lowered, especially for galaxies which present star
formation (TOs and SFN). It is noteworthy that while radio AGN and passive
galaxies show the same general trend with respect to density, they show opposite
trends with respect to the tidal and PCA2 parameters. The relations with respect
to PCA3r are usually non-linear (see Fig.~\ref{fig:prev}); these type of
non-linear trends were also observed by \citet{Pimbblet2012arxiv} for the
fraction of optical AGN with respect to the distance to the center of a cluster.
This non-linearity weakens the trends found using only two bins, making it
difficult to draw strong conclusions. Nevertheless, some results are apparent. A
lower fraction of optical AGN is generally found for lower values of PCA3r (high
density smaller systems, e.g. compact groups). There is also a possible decrease
of the fraction of LINERS and TOs towards higher values of PCA3r (e.g. cluster
outskirts).

\subsection{Cochran-Mantel-Haenzsel test}

The Cochran-Mantel-Haenzsel test \citep[CMH;][]{Cochran1954,Mantel1959} can be
used to study the strength of the association between two bi-valued variables
(odds ratio) and its significance after taking into account the effect of the
possible confounding factors. In our case, we corrected for the effect of the
mass. The first variable was the ``exposure'' to the environmental parameter and
the second variable was the effect of this exposure on the triggering of certain
type of nuclear activity. We discern the ``exposure'' to the environmental
parameter by separating galaxies above and below the median value for a given
environmental or PCA parameter. In the case of the cluster richness, more than a
half of the galaxies have an assigned $n$ of 1. Hence, the two selected groups
were galaxies with $n = 1$ and galaxies with $n \geq 2$. 

We tested if the prevalence of galaxies with certain type of activity is
independent of the environment or interaction traced by the belonging to one of
the two groups described before (galaxies with high or low values of the
environmental parameter or PCA). We obtained the CMH ratio and its significance.
The CMH ratio is the relative probability between the groups with high and low
environmental parameters. A ratio compatible with 1 means that the prevalence of
nuclear activity is independent of the environmental or PCA parameter or that a
dependence cannot be found with the data used. A significant ratio (defined by
the $p$-value), higher or lower than 1, indicates that this environmental or PCA
parameter is increasing or decreasing the chance for a galaxy to harbour this
type of nuclear activity.

\begin{table*}
 \centering
 \begin{minipage}{140mm}
  \caption{Statistical study results from the CMH test. For each cell of the
table the first row shows the CMH common odds ratio and its error and the second
row shows the statistical significance or $p$-value (i.e., the probability of
the trend occurring by chance) measured by the CMH test. The hypothesis tested
was that the observed nuclear activity type is independent of the environmental
or PCA parameter. The typeface of the $p$-value depends on its value: bold if $p
< 0.01$; bold italics if $0.01 \leq p < 0.05$; regular if $0.05 \leq p < 0.1$;
and italics if $p \geq 0.1$}
  \label{table:stats}
   \begin{tabular}{l|cccccc}
\hline
 & Density & Tidal & log(n) & PCA1 & PCA2 & PCA3r\\
\hline
\hline
Passive & $1.358 \pm 0.061$ & $0.950 \pm 0.043$ & $1.248 \pm 0.077$ & $1.143
\pm
0.052$ & $0.745 \pm 0.034$ & $0.989 \pm 0.055$\\ 
  & \textbf{0.0000} & \textit{\textbf{0.0296}} & \textbf{0.0000} &
\textbf{0.0000} & \textbf{0.0000} & \textit{0.7025}\\ 
~~LINER & $ 0.77 \pm  0.11$ & $ 0.84 \pm  0.13$ & $ 0.87 \pm  0.14$ & $ 0.80
\pm 
0.12$ & $ 1.08 \pm  0.16$ & $ 1.18 \pm  0.19$\\ 
  & \textbf{0.0012} & \textit{\textbf{0.0353}} & \textit{0.1248} &
\textbf{0.0053} & \textit{0.3375} & 0.0642\\ 
~~Seyfert & $0.801 \pm 0.054$ & $0.898 \pm 0.061$ & $0.783 \pm 0.070$ & $0.807
\pm
0.055$ & $1.118 \pm 0.076$ & $0.914 \pm 0.076$\\ 
  & \textbf{0.0000} & \textbf{0.0029} & \textbf{0.0000} & \textbf{0.0000} &
\textbf{0.0018} & \textit{\textbf{0.0406}}\\ 
~~TO & $0.807 \pm 0.076$ & $ 1.08 \pm  0.10$ & $ 0.84 \pm  0.10$ & $0.906 \pm
0.086$ & $ 1.35 \pm  0.13$ & $ 0.97 \pm  0.11$\\ 
  & \textbf{0.0000} & \textit{0.1568} & \textbf{0.0097} & 0.0510 &
\textbf{0.0000} & \textit{0.6727}\\ 
SFN & $0.608 \pm 0.051$ & $ 1.39 \pm  0.11$ & $0.787 \pm 0.100$ & $0.999 \pm
0.082$ & $ 1.84 \pm  0.15$ & $ 1.16 \pm  0.13$\\ 
  & \textbf{0.0000} & \textbf{0.0000} & \textbf{0.0005} & \textit{0.9783} &
\textbf{0.0000} & \textbf{0.0085}\\ 
Optical & $0.797 \pm 0.042$ & $0.939 \pm 0.050$ & $0.808 \pm 0.056$ & $0.831
\pm
0.044$ & $1.181 \pm 0.062$ & $0.964 \pm 0.061$\\ 
AGN & \textbf{0.0000} & \textit{\textbf{0.0253}} & \textbf{0.0000} &
\textbf{0.0000} & \textbf{0.0000} & \textit{0.2790}\\
\hline
Radio & $ 1.55 \pm  0.21$ & $ 2.77 \pm  0.52$ & $ 2.00 \pm  0.28$ & $ 2.13 \pm 
0.36$ & $ 1.31 \pm  0.18$ & $ 1.20 \pm  0.16$\\ 
AGN & \textbf{0.0000} & \textbf{0.0000} & \textbf{0.0000} & \textbf{0.0000} &
\textbf{0.0003} & \textit{\textbf{0.0113}}\\ 
~~LERG & $ 1.63 \pm  0.23$ & $ 2.96 \pm  0.59$ & $ 2.07 \pm  0.30$ & $ 2.21 \pm 
0.38$ & $ 1.30 \pm  0.18$ & $ 1.19 \pm  0.16$\\ 
  & \textbf{0.0000} & \textbf{0.0000} & \textbf{0.0000} & \textbf{0.0000} &
\textbf{0.0004} & \textit{\textbf{0.0185}}\\ 
~~HERG & $ 0.55 \pm  0.27$ & $ 1.01 \pm  0.52$ & $ 1.07 \pm  0.56$ & $ 1.09 \pm 
0.57$ & $ 1.47 \pm  0.75$ & $ 1.81 \pm  0.97$\\ 
  & 0.0720 & \textit{0.9881} & \textit{0.8496} & \textit{0.8141} &
\textit{0.2897} & \textit{0.1248}\\
\hline
\end{tabular}
\end{minipage}
\end{table*}

\begin{figure*}
\centering
 \includegraphics[width=15cm]{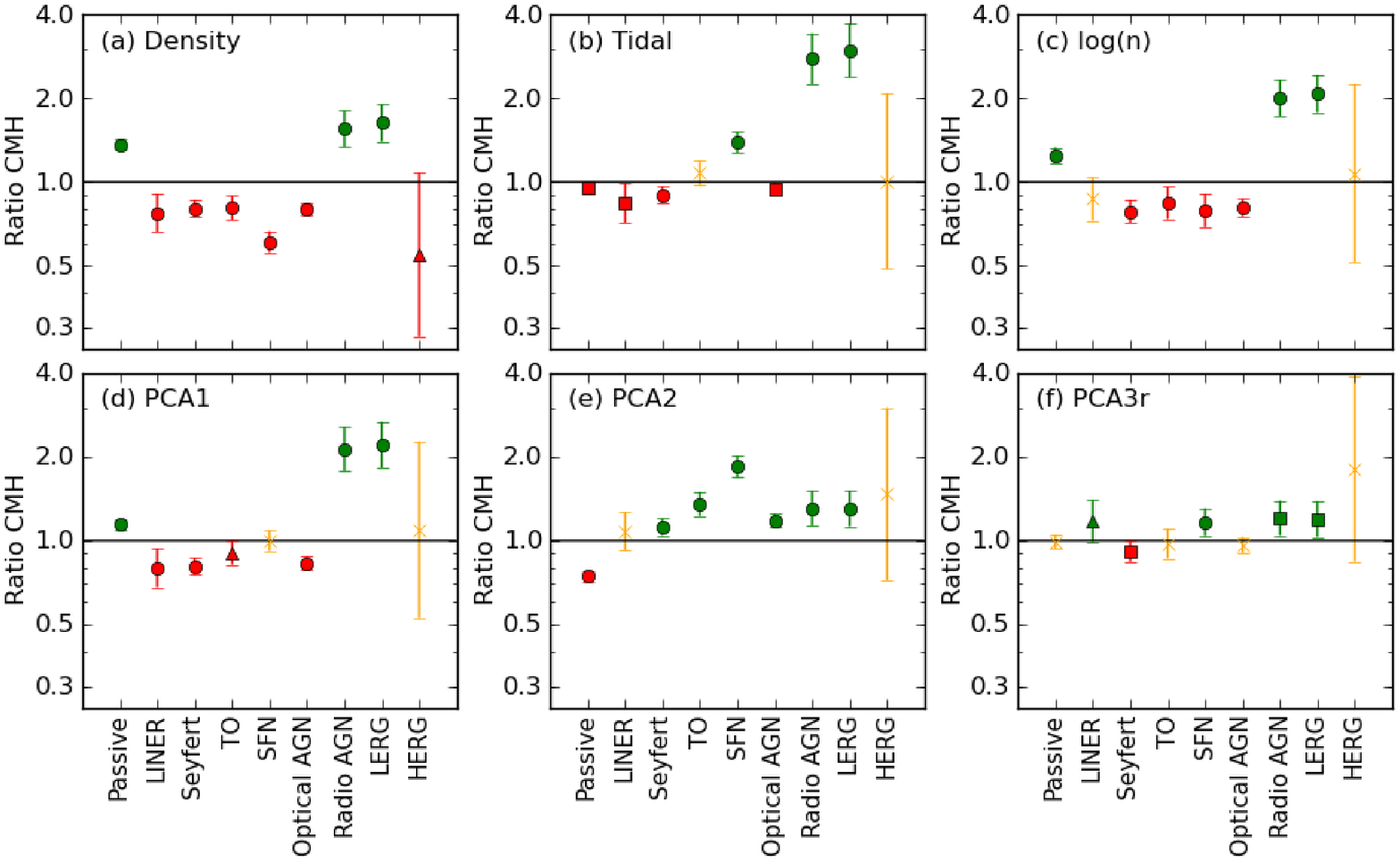}
  \caption{Cochran-Mantel-Haenzsel ratios for the different types of nuclear
activity with respect to the environmental parameters. The error bars mark the
95 per cent confidence interval. The shape of the symbol indicates the
confidence level of the hypothesis tested (the probability of harbouring this
type of activity is not affected by the environmental parameter) measured by the
\textit{p}-value: circle if $p < 0.01$; square if $0.01 \leq p < 0.05$; triangle
if $0.05 \leq p < 0.1$; and cross if $p \geq 0.1$.}
  \label{fig:ratios}
\end{figure*}

The results of the CMH statistical test are shown in Table~\ref{table:stats} and
in Fig.~\ref{fig:ratios} they are shown in a way that allows a quick visual
comparison between them. We added in this case a separation of radio AGN into
HERG and LERG types. An increase in the local density produces a decrease in the
prevalence of optical AGN (LINER, Seyferts, TOs) and SFN. The effect is the
opposite for radio AGN and passive galaxies. When we separate radio AGN in HERG
and LERG, they present opposite trends with respect to the local density: LERGs
show a clear increase with density while HERGs appear to show a decrease like
optical AGN (though uncertainties are large due to the small sample size). The
effect of the tidal parameter is not clear for passive galaxies and TOs (as
expected from the non-linear dependence in Fig.~\ref{fig:prev}). A slight
decrease of the fraction of LINERs and Seyferts with increasing tidal parameter
is found while an increase is found for SFN and radio AGN. This might be caused
by the non-linear relation due to the effect of high density galaxies explained
before. The trends with respect to the cluster richness are similar to the
trends with density. The effects that depend on PCA1 are very similar to the
ones driven by density except for SFN. Probably the decreasing prevalence of SFN
with density is compensated by the increasing prevalence with the tidal
estimator to make the PCA1 relation flat. In the case of increasing PCA2, there
is an increase in the AGN fraction for Seyferts, TOs, SFN and radio AGN and a
clear decrease for passive galaxies. The trends for HERGs and LERGs are
compatible here. PCA2 reveals much stronger positive trends for optical AGN than
tidal does, due to removal of density selection effects, i.e. using PCA we show
the importance of the one-on-one interactions. The radio AGN trend with PCA2 is
much weaker than that with the tidal parameter, indicating that much of what is
seen with respect to the tidal parameter is just a density effect (which gets
stronger going from density to PCA1). The trends with PCA3r generally tend to be
weaker and more variable for optical AGN types, as expected from the
non-linearities shown in Fig.~\ref{fig:prev}. However, radio AGN and SFN have
clearly increasing fractions with an increasing PCA3r.


\section{Discussion}
\label{sec:dc}

\subsection{Environment and interactions}

Of the parameters studied, mass is confirmed to be the main driving factor for
the triggering of an AGN, especially for radio AGN
\citep{Best2005b,Kauffmann2004} as shown in Fig~\ref{fig:mass_slices}. However,
both local density and interactions also play a significant role on the
prevalence of nuclear activity. Fig.~\ref{fig:ratios} and
Table~\ref{table:stats} show the clear relations between the prevalence of the
different types of nuclear activity with the environmental parameters. The
relations obtained show the clearly different and even opposite trends found for
the AGN fraction depending on the environmental or PCA parameter used and the
type of AGN studied. These differences may have led to previous discrepancies
found in the literature.

The fraction of optical AGN decreases toward higher values of the local density,
PCA1 and the cluster richness, with the prevalence of AGN in the densest local
environments being a factor 2--3 lower (at fixed mass) than that in sparser
environments. The similarity between the trends with respect to local density
and cluster richness reflects that both parameter are closely related, as shown
in the panel \textbf{c} of Fig.~\ref{fig:params}, with galaxies with higher $n$
located at higher densities. In rich environments there are several mechanisms
that can remove or change the properties of the cold gas needed to fuel the star
formation or AGN from galaxies: the stripping of diffuse gas around galaxies or
strangulation, ram-pressure stripping and galaxy harassment \citep[see][ and
references therein]{vonderLinden2010}. The stripping of cold gas and its warming
can explain the decrease of the prevalence of optical activity (AGN and star
formation) found in denser and richer environments.

On the other hand, one-on-one interactions, traced by PCA2, produce an
enhancement on the prevalence of SFN, TOs, Seyfert and radio AGN and a decrease
of the fraction of passive galaxies. This agrees with an scenario where galaxy
interactions can fuel gas to the central region of the galaxy
\citep{Shlosman1990,Barnes1991,Haan2009,Liu2011}. However, notice that this is
not the only mechanism that triggers nuclear activity
\citep[e.g.][]{Silverman2011,Sabater2012}; the AGN prevalence only increases by
a factor $\sim 2$ between galaxies with no nearby neighbours to those undergoing
the strongest interactions. For optically selected nuclear activity, the values
of the CMH ratios with respect to PCA2 (see panel \textit{e} in
Fig.~\ref{fig:ratios}) decrease in this order: SFN, TOs, Seyfert, LINERs and
passive galaxies. This relation can be compatible with the existence of a time
delay between the tidal interaction and the setting of the different types of
activity with an interaction inducing first star formation, then a Seyfert-like
AGN, and then a LINER, before returning to a passive state
\citep[e.g.][]{Li2008,Darg2010,Wild2010}.

The slightly decreasing trend of AGN prevalence found for optical AGN (both
LINERs and Seyferts) with respect to the tidal parameter (see panel \textbf{b}
in Fig.~\ref{fig:ratios}) may be produced by an overestimation of the tidal
parameter in regions with high density of galaxies due to the use of projected
distances (as dicussed earlier). Another explanation is that galaxies with a
high tidal parameter located in dense environments lack the supply of gas
required to fuel their nuclear activity despite their ongoing one-on-one
interaction (see next Section). When this effect is corrected for using the
correlation found with the PCA by the use of PCA2, the increase of the
prevalence of all types of active galaxies with PCA2 becomes conspicuous. This
suggests that PCA2 may trace better one-on-one interactions than the uncorrected
tidal estimator, or at least, it may better trace the sort of one-on-one
interactions that can trigger nuclear activity (i.e. those that fuel gas into
the nucleus).

Interactions and environmental density are complementary factors that must be
taken into account to get a complete view of the external phenomena that affect
the triggering of an AGN. The trends found can be explained if the different
parameters trace different physical processes that affect the triggering of
nuclear activity.

\subsection{AGN types}

Optical AGN and HERG are thought to be fuelled in different ways than the
prevalent population of radio AGN \citep[LERG;][]{Best2012}. The first type of
AGN are fuelled at relatively high Eddington rates in radiatively efficient
accretion discs while the second type is fuelled at relatively low Eddington
rates in radiatively inefficient accreting flows. While the first type requires
a plentiful supply of cold gas, the lower accretion-rate requirements of the
LERGs could be achieved by accreting warm gas located in the haloes surrounding
the galaxy, cluster or group. 

The local density parameter can trace the content and physical condition of the
gas. The lack of cold gas in clusters of galaxies in the local Universe is well
known \citep[e.g.][]{Giovanelli1985}, and the gas that can be found at higher
densities is warmer \citep[see][]{Cattaneo2009}. If a plentiful supply of cold
gas is required to fuel optical AGN and SFN, then these types of activity will
be suppressed in dense environments but enhanced by one-on-one interactions, as
discussed above. On the other hand, the accretion of warm gas in would explain
the higher prevalence of radio AGN of LERG type in denser environments. The
result that LERGs are enhanced by one-on-one interactions, however, argues
against the simplest models in which all LERGs are fuelled from the warm gas
haloes around massive galaxies, groups or clusters
\citep[e.g.][]{Hardcastle2007}. It suggest that some LERGs are fuelled by other
mechanisms, and supports arguments that the distinction between radiatively
efficient and radiatively inefficient AGN is more likely to be a function of
Eddinton-scaled accretion rate \citep[e.g.][]{Merloni2008,Best2012} than the
origin of the fuelling gas. In this picture, most galaxy interactions provide
gas to the nucleus at a sufficiently high rate that a radiatively efficient
(optical) AGN is produced, but some interactions (or at some times during an
interaction) the gas flow to the nucleus is slower and results in a LERG. These
LERGs may be distinguishable from the cooling-fuelled LERGs in terms of their
colours \citep[e.g.][]{Janssen2012}.

The low numbers of HERGs make difficult to draw statistically significant
results from them, but their prevalence is statistically higher at lower than at
higher local densities (see panel \textit{a} in Fig.~\ref{fig:ratios}). This is
opposite to the trend found for LERGs but the same as that for optical AGN.
Indeed, all the environmental trends for HERGs are more similar to optical AGN
than to the rest of radio AGN (LERGs). This suggests that the physical processes
that trigger HERGs are similar to that of optical AGN. Despite the similar
powering mechanisms and environment trends, it is important to note that HERGs
do show significant differences with respect to Seyferts, most particularly that
they are preferentially found at higher stellar masses (median values for the
whole sample: $10^{10.88}$ vs. $10^{10.56}$ M$_{\odot}$; reduced sample:
$10^{10.95}$ vs. $10^{10.74}$ M$_{\odot}$).

The mechanisms that power LINER galaxies are a matter of controversy
\citep{Kewley2006}. Some LINERs are certainly powered by accretion
\citep[e.g.][]{Omaira2006}, but there are strong arguments that a fraction of
them could be powered by low-mass evolved stars
\citep[e.g.][]{Cid-Fernandes2011}. In our analysis we found that the LINERs show
very similar trends to those of the rest of the optical AGN population (although
some differences can be observed in PCA3r). This may be caused by the limiting
luminosity that we apply to $L_{[\ion{O}{iii}]}$, which will select only the
most luminous LINERs which are more likely to be powered by AGN
\citep{Cid-Fernandes2011}; these may behave differently from low-luminosity AGN
\citep{Schawinski2010}. This may cause the trends of our LINERs to be similar in
general to those of Seyfert galaxies. 

\subsection{Further considerations}

In \citet{Sabater2012}, no evidence of a difference was found between the
prevalence of AGN in isolated galaxies \citep[AMIGA sample of isolated
galaxies;][]{Verdes-Montenegro2005} or in a sample of compact groups
\citep[][]{Martinez2010} at a fixed stellar mass and morphology. This result can
be explained if the increment expected in compact groups due to higher
one-on-one interactions is broadly compensated by their higher local density and
deficiency of cold gas \citep{Verdes-Montenegro2001}. However, an interpretation
of the differences found for star formation in the literature
\citep[e.g.][]{Sulentic2001,Durbala2008b,Martinez-Badenes2012} is not
straightforward.

There are additional considerations that could be taken into account to further
advance this study. Firstly, morphology could play an important role on nuclear
activity \citep{Hwang2012}, and its effect as a confounding factor could also be
estimated. Secondly, lower emission line luminosity systems could be probed to
investigate the more typical LINER population. To allow the discrimination
between genuine AGN-powered LINERs and retired galaxies whose emission resembles
that of weak AGN, alternative diagnostic methods that take into account the
differences between these types of weak AGN would have to be considered
\citep{Buttiglione2010,Cid-Fernandes2011,Sabater2012}. Thirdly, it could be
investigated whether different sub-populations of LERGs (e.g. by colour or
luminosity) show the same environmental and interaction dependencies. This study
will be extended in the future to clarify those points. 

\section{Summary and conclusions}
\label{sec:conc}

We presented a study of the prevalence of optical and radio nuclear activity
with respect to the environment and interactions in a sample of SDSS galaxies.
We aimed to determine the effect of these external factors on the triggering of
nuclear activity but taking into account the effect of possible confounding
factors like the galaxy mass. 

The study is based on a sample of $\sim$270000 galaxies with a redshift between
0.03 and 0.1 drawn from the main spectroscopic sample of the Sloan Digital Sky
Survey (SDSS). Additional data like the stellar mass and the optical AGN
classification were obtained from the MPA-JHU catalogue. The radio AGN
classification is described in \citet{Best2012}. We defined a local density
parameter and a tidal forces estimator and used a cluster richness estimator
from the literature \citep{Tago2010} to trace different aspects of environment
and interaction. A principal component analysis was applied to the environmental
parameters to consider and remove the possible correlations between them. A
stratified statistical study, which removed the effect of the galaxy mass, was
applied in order to find the unbiased relation of the prevalence of the
different types of nuclear activity with the different PCA and environmental
parameters. The relations were quantified and their statistical significance
obtained using the Cochran-Mantel-Haenzsel test.

From this study we found that
\begin{enumerate}
  \item There is a strong dependence of the prevalence of nuclear activity with
density and interactions after taking into account the effect of mass. However,
it is fundamental to distinguish and consider both  local environment and
one-on-one interactions because they can produce different and even opposite
effects.
  \item We found a decrease of optical nuclear activity, including star
formation, towards denser environments. This effect is probably driven by the
lack of a cold gas supply at higher densities. 
  \item On the other hand, an increase of the prevalence of radio nuclear
activity is found towards higher densities. This can be explained by the warmer
gas present at higher densities that is accreted at low rates in a radiatively
inefficient manner, triggering \textit{typical} low-luminosity radio AGN or
LERGs.
  \item There is an increase of the prevalence of SFN, optical and radio AGN
with one-on-one interactions. The dependence for the different types of optical
activity with respect to PCA2 traced by the CMH ratio follows this order: SFN
(1.84), TOs (1.35), Seyferts (1.12), LINERs (1.08) and, finally, passive
galaxies (0.75); this is in accordance with the time sequence that has been
argued for the different activity types.
  \item The trend for HERGs with local density is statistically different to
that of the rest of radio AGN (LERG) and similar to that of optical AGN. This
adds evidence to the link of HERG with high excitation AGN, like optical and
X-ray AGN, that needs a supply of cold gas to be powered.
\end{enumerate}

Our results agree with a scenario where the type of AGN triggered is related to
the presence and physical condition of the gas supply of the AGN, which in turn
depends on both large-scale environment and local interactions. High excitation
AGN (optical AGN and HERGs) and SFN require the presence of a plentiful cold gas
supply to the nuclear regions. This gas is less likely to be present in denser
environments where galaxies tend to be gas poor due to stripping or harassment.
The prevalence of AGN is enhanced in the presence of one-on-one interactions
which can funnel gas to the nuclear regions. These two effects may work in
competition against each other (one-on-one interactions are more likely in
denser environments) and hence, the consideration of different aspects of the
environment is fundamental to obtain a complete understanding of the mechanisms
that may trigger the different types of AGN. In contrast, low-luminosity radio
AGN of LERG type require relatively low gas accretion rates. This can be
supplied by the cooling of gas from the hot haloes of groups and clusters
accounting for the strong increase in the prevalence of LERGs with local
density. However, the presence of a residual trend with one-on-one interactions
after density effects are removed (PCA2) suggest that warm gas cooling is not
the only mechanism by which LERGs can be triggered.

%

\section*{Acknowledgments}
We acknowledge the useful comments of the anonymous referee. JS and MAF were
partially supported by Grant AYA2011-30491-C02-01, co-financed by MICINN and
FEDER funds, and the Junta de Andaluc\'{\i}a (Spain) grants P08-FQM-4205 and
TIC-114. Funding for the SDSS and SDSS-II has been provided by the Alfred P.
Sloan Foundation, the Participating Institutions, the National Science
Foundation, the U.S. Department of Energy, the National Aeronautics and Space
Administration, the Japanese Monbukagakusho, the Max Planck Society, and the
Higher Education Funding Council for England. The SDSS Web Site is
http://www.sdss.org/. The SDSS is managed by the Astrophysical Research
Consortium for the Participating Institutions. The Participating Institutions
are the American Museum of Natural History, Astrophysical Institute Potsdam,
University of Basel, University of Cambridge, Case Western Reserve University,
University of Chicago, Drexel University, Fermilab, the Institute for Advanced
Study, the Japan Participation Group, Johns Hopkins University, the Joint
Institute for Nuclear Astrophysics, the Kavli Institute for Particle
Astrophysics and Cosmology, the Korean Scientist Group, the Chinese Academy of
Sciences (LAMOST), Los Alamos National Laboratory, the Max-Planck-Institute for
Astronomy (MPIA), the Max-Planck-Institute for Astrophysics (MPA), New Mexico
State University, Ohio State University, University of Pittsburgh, University of
Portsmouth, Princeton University, the United States Naval Observatory, and the
University of Washington.

\bibliographystyle{mn2e_warrick}
\bibliography{databasesol}

\label{lastpage}

\end{document}